\documentclass[10pt,letterpaper,twocolumn]{article} 

\usepackage{ol2,units}
\usepackage{amsmath}

\begin{document}

\twocolumn[ 

\title{Continuous phase stabilization and active interferometer control using two modes}


\author{Gregor Jotzu, Tim J. Bartley$^{*}$, H. B. Coldenstrodt-Ronge, Brian J. Smith, and I. A. Walmsley}

\address{
University of Oxford, Clarendon Laboratory, Parks Road, Oxford, OX1 3PU, United Kingdom \\
$^*$Corresponding author: {\rm{t.bartley1@physics.ox.ac.uk}}}
\begin{abstract}
We present a computer-based active interferometer stabilization method that can be set to an arbitrary phase difference and does not rely on modulation of the interfering beams. The scheme utilizes two orthogonal modes propagating through the interferometer with a constant phase difference between them to extract a common phase and generate a linear feedback signal. Switching times of {\unit[50]{ms}} over a range of $0$ to $6\pi$ radians at {\unit[632.8]{nm}} are experimentally demonstrated. The phase can be stabilized up to several days to within $\pm 3^{\circ}$.
\end{abstract}


] 

The ability to continuously adjust and stabilize the optical phase difference between two arms of an interferometer is of great importance to a wide range of applications including homodyne detection for quantum state tomography~\cite{smithey93}, phase-shift keying in optical telecommunications~\cite{gnauck05}, phase-shifting interferometry \cite{creath98}, ultrafast pump-probe spectroscopy~\cite{zhang05, branderhorst08, brinks10}, interference-based optical lattices~\cite{sebby-strabley06} and near-field scanning microscopy~\cite{gersen05}. Previous approaches typically relied on fringe-lock methods, in which a useful error signal can only be produced near integer multiples of $\pi/2$. These methods are prone to fringe skipping, where noise causes the phase lock circuit to hop to a neighboring interference fringe, a half wavelength away. In many applications the ability to continuously adjust the phase difference to an arbitrary value is required, which cannot be achieved with fringe-lock methods. Continuous phase locking has been previously accomplished in two ways. An arbitrary phase can be locked by modulation of one of the interfering beams and detection at the fundamental and second harmonic of the modulation frequency~\cite{freschi95}. This method, which produces a sinusoidal error signal, is still susceptible to fringe skipping. Furthermore, the phase modulation requires specialized signal analysis and is undesirable when signal acquisition rates faster than the modulation frequency are necessary. 
A recent technique that utilizes tilting of the beam in one arm of the interferometer and spatially resolved measurements at the output to create a linear error signal was introduced~\cite{krishnamanchari06}. However, this scheme requires precise alignment and stabilization of photodetector positioning, and suffers from chromatic aberrations of the glass wedge used for the tilt.

In this Letter, we present a general approach to phase control capable of stabilization to an arbitrary phase setting by producing a linear error signal in the phase. The basic principle lies in utilizing two distinct optical modes passing through the interferometer with non-identical optical path length differences resulting in two phase offsets. The modes could consist of different transverse-spatial, temporal, frequency or polarization modes depending upon the nature of the experiment and noise involved. Indeed, this mode multiplexing approach can be seen as the key physical principle behind previous schemes to continuous phase stabilization, which utilized frequency~\cite{freschi95}, and spatial modes~\cite{krishnamanchari06}. Both modes experience a common phase shift $\phi$, and one mode undergoes an additional, but constant phase offset $\delta$. Monitoring both modes at the output of the interferometer, with knowledge of the offset phase difference $\delta$, enables an accurate linear estimation of the phase common to both, and thus stable feedback control over a continuous range of $\phi$. 
\begin{figure}[h]
\centering
\includegraphics[width=0.95\linewidth]{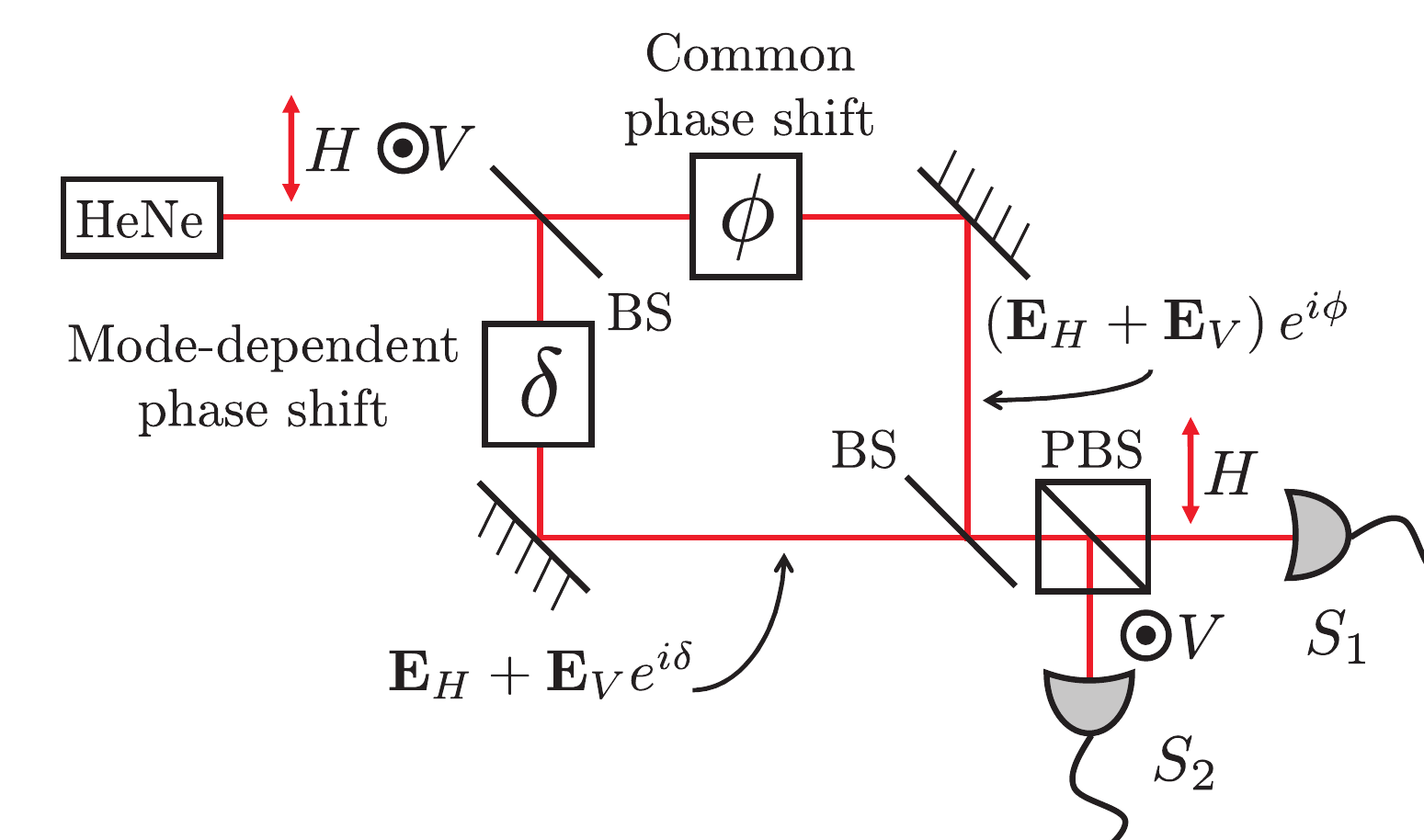}
\caption{\label{setup}Setup of our Mach-Zehnder interferometer (MZI) with two polarization modes. Both horizontal ($H$) and vertical ($V$) modes are subject to a controllable common phase $\phi$, while one mode, $V$, picks up an additional phase $\delta$ in the lower branch of the MZI. The interference signals $S_{1}$ and $S_2$ of each mode are measured independently, from which the phase $\phi$ is calculated by software.}
\end{figure}

As a concrete example, consider two polarization modes as depicted in Fig.~\ref{setup}. A reference laser polarized at $45^{\circ}$ with respect to the horizontal is launched into a balanced Mach-Zehnder interferometer (MZI), constructed with non-polarizing $50$:$50$ beam splitters (BS). Both polarizations traverse the upper path containing a controllable phase $\phi$. A waveplate in the lower path of the MZI produces a constant relative phase shift $\delta$ between the horizontal and vertical polarization modes. The polarization modes from one output port of the MZI are split by a polarizing beam splitter (PBS), and directed to photodiode detectors. This scheme is easily generalized to other modes, e.g. two transversely displaced co-propagating laser beams.

The output currents from the photodiodes are converted into voltages across resistors and monitored by a computer using a fast analog-to-digital converter (ADC). The expected signal voltages have the form
\begin{equation}
S_1=A_1\cos(\phi)+ B_{1}, \quad S_2=A_2\cos(\phi+\delta)+ B_{2}.
\label{voltages}
\end{equation}
Here $A_{1(2)}$ and $B_{1(2)}$, which depend on the input intensity, interference visibility, detector gain, background light and electronic noise levels, are calibrated prior to stabilization. These equations can be combined to estimate the common phase value
\begin{equation}
{\phi} = U\left\{\tan^{-1}{\left[\frac{\cos(\delta)}{\sin(\delta)} - \frac{A_1(S_2-B_{2})}{A_2(S_1-B_{1})\sin(\delta)}\right]} \right\},
\label{phase}
\end{equation}
where the $U$ is the phase unwrapping function. This estimate works accurately as long as the phase does not change by more than $\pi$, corresponding to a half wavelength path difference, within a sampling period. If $\phi_0$ is the desired phase setting at which the interferometer is to be locked, the estimated phase in Eq. (\ref{phase}) can be used to implement a linear error signal to drive the feedback control of $\phi$
\begin{equation}
err = - f (\phi - \phi_0),
\label{error}
\end{equation}
where $f$ is a constant feedback gain parameter.

To experimentally test this phase stabilization technique we used a Helium-Neon (HeNe) reference laser (JDS Uniphase 1125P, $5$~mW average power) polarized at $45^{\circ}$ with respect to the horizontal, as shown in Fig. \ref{setup}. The phase $\phi$ in the upper arm of the MZI is set by a computer controlled piezo-electric translation (PZT) stage (Thorlabs NF5DP20/M) delay line. A quarter wave plate (QWP) with vertically aligned slow axis sets the constant relative phase $\delta \approx 90^{\circ}$ in the lower MZI arm. Note that a wide range of $\delta$ can be tolerated, so that a multi-order wave plate or arbitrary birefringent medium could potentially be used. A PBS at one output of the MZI splits the horizontal and vertical polarization modes, which are subsequently focused onto fast photodiodes (Thorlabs DET10A/M). The voltages produced by the photodiodes are monitored using a fast 16-bit analog-to-digital / digital-to-analog converter (AD/DAC) (National Instruments USB-6221 BNC) connected to a personal computer (PC) at a rate of $10$ kHz. A software program calculates the error signal in Eq. (\ref{error}), which is sent to the PZT stage using the AD/DAC.

\begin{figure*}[th!]
\centering
\includegraphics[width=0.95\linewidth]{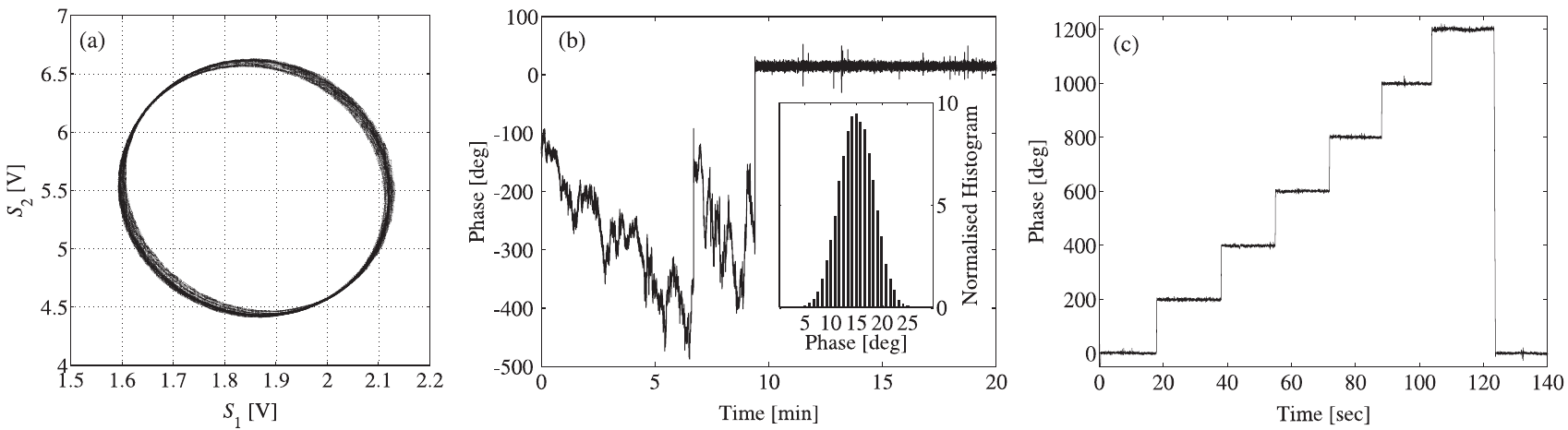}
\caption{\label{fig:results}(a) Parametric plot of $\left(S_1\left(\phi\right),S_2\left(\phi\right)\right)$ when scanning the phase $\phi$ yields the calibration ellipse. By fitting the data, the values of the constants in Eq.~(\ref{phase}) are determined. (b) After allowing the phase to fluctuate naturally for 10 minutes, the stabilizer is switched on to lock the phase at $15^\circ$. The root-mean-squared error in the locked phase is $3^\circ$, which can be maintained for over 24 hours. (c) The phase of the stabilizer can be arbitrarily set to any value within the range of the computer controlled delay stage. Here, we show phase locking to various phases from 0 to $6\pi$~radians.}
\end{figure*}

To implement the stabilization scheme the parameters $A_{1(2)}$, $B_{1(2)}$ and $\delta$ must first be calibrated. This is done by scanning the phase $\phi$, and collecting the detector voltages. A parametric plot of $\left(S_2\left(\phi\right),S_1\left(\phi\right)\right)$ as the phase $\phi$ is scanned sweeps out a rotated and displaced ellipse, as shown in Fig.~\ref{fig:results} (a). Fitting this ellipse to Eq.~(\ref{voltages}) directly yields the calibration parameters to be used in the phase stabilization protocol. To obtain more accurate values of the feedback parameters, we repeated this process multiple times. As long as $\delta$ differs sufficiently from $0$ and $\pi$, i.e. the ellipse does not become a line, successful stabilization can be accomplished. Roughly speaking, the phase $\phi$ can be stabilized to within 
\begin{equation}
\Delta \phi \geq \frac{\Delta P}{\bar{P}}\frac{1+\sqrt{2}}{2V | \sin(\delta/2) |},
\label{uncertain}
\end{equation}
where $\bar{P}$ and $\Delta P$ are the average and standard deviation in the reference laser power, $V$ is the fringe visibility for the two modes (assumed equal), and $\delta$ is the phase offset between the two modes. This can be derived from assuming the signal voltages are given by $S_1 = gP (1 + V \cos (\phi) )/2$ and $S_2 = gP (1 + V \cos (\phi+\delta) )/2$, where $g$ is the detector gain, assumed to be equal for both detectors. This result puts quantitative bounds on the intensity fluctuations, fringe visibility, and offset phase that can be tolerated to yield phase stabilization within $\Delta \phi$.

Once calibrated, the interferometer can be stabilized at a desired setting $\phi_{0}$, by continuous monitoring of the phase $\phi$ and feedback to the control PZT stage on a ms timescale. The feedback voltage applied to the PZT, derived from Eq.~(\ref{error}), was $V_f = - f \left\langle \phi - \phi_{0} \right\rangle$, where the average was taken over previous measurements. The best stability was achieved with a feedback gain of $f$~=~\unit[3.65]{mV/degree}. The optimal feedback function, that is, the relationship between error signal and the feedback control voltage, will generally depend on the details of the experiment~\cite{bechhoefer05}.

The performance of our active feedback stabilization scheme was limited primarily by power fluctuations of the reference laser and the feedback response time (i.e. the time delay between estimating the phase and the implementation of the feedback signal voltage). Power fluctuations can be eliminated by monitoring the input laser power and normalizing the output data, or use of a more stable laser system. The time lag was mainly due to the PC processing time and PZT response time, which varied between $4$~ms and $7$~ms. This implies high-frequency noise could not be compensated. This issue could be improved using faster AD/DAC electronics and lower-level programming to implement the phase estimation and feedback.
The results shown in Fig. \ref{fig:results} (b) and (c) were achieved with calibration parameters $A_1=0.260$, 
$A_2=1.083$,  
$B_{1}=1.859$,  
$B_{2}=5.511$, 
and $\delta=93.5^{\circ}$.  
At what would be an unstable phase setting for fringe lock methods, $\phi_0 = 15^{\circ}$, we are able to stabilize to within a standard deviation of $\Delta \phi \approx 3^{\circ}$. This corresponds to length variations of approximately \unit[5]{nm} between the two \unit[3.5]{m} long arms of the interferometer. Given we have power fluctuations $\Delta P/\bar{P}$ of 0.6\%, Eq.~\ref{uncertain} yields phase fluctuations of $2.9^\circ$, which is in good agreement. 

This stability can be maintained for over 24 hours. Furthermore, the desired phase setting $\phi_{0}$ can be changed in real time as shown in Fig.~\ref{fig:results} (c). Changing the phase setting $\phi_{0}$ by more than $6\pi$ could be achieved with switching time near \unit[50]{ms}, which is primarily limited by the feedback response time. The current scheme thus allows one to set and stabilize an interferometer to any optical path length difference with a precision of a few nanometers and a range limited only by the movement of the PZT stage (about \unit[35]{$\mu$m}), and ultimately the coherence length of the reference laser.

In conclusion, we have shown a general approach to interferometric phase control capable of locking to any chosen phase value. The key element in this scheme is the use of two orthogonal modes with known, fixed phase offset to obtain a precise estimate of the interferometer phase for arbitrary path length difference. This enables highly accurate feedback control of the system. Depending upon the nature of the interferometer to be stabilized, in particular, considering the main sources of phase noise, choice of what degree of freedom to utilize can be made to optimize the stabilization scheme. This general approach to phase stabilization allows control under diverse noise conditions. Polarization modes are extremely useful when there is little change in birefringence between the two interferometer arms as demonstrated here. Frequency modes could be used in optical fiber based interferometers where the dispersion is known. Improvements to the feedback electronics will likely lead to faster switching times and increased stability against higher-frequency noise. To increase the precision of this scheme requires increased stability or continuous monitoring of the reference laser, as eluded to in Eq.~(\ref{uncertain}). Furthermore, implementing a more sophisticated feedback signal, e.g. using a PID control algorithm~\cite{bechhoefer05}, should lead to improved performance. The techniques developed here should be of use for a wide range of applications due to the applicability of the stabilization scheme to all optical degrees of freedom.

This research was supported by the EU Integrated Project Q-ESSENCE through the European Community's Seventh Framework Programme FP7/2007-2013 under grant agreement no. 248095. The authors would like to thank Gaia Donati for assistance with the data analysis.


\end{document}